NOTE DE RECHERCHE
# PRESENTATION DU NOUVEL ACCORD DE BALE SUR LES FONDS PROPRES.

**Hamza FEKIR**[1]

**Résumé :**

Afin de s'adapter à la libéralisation de la sphère financière entamée dans les années 80, marquée notamment par la fin de l'encadrement de crédit, la disparition des différentes formes de protection de l'Etat dont bénéficiaient les banques, et la privatisation de la quasi-totalité des établissements en Europe, la réglementation bancaire a évolué vers une approche prudentielle, perçue comme le seul mode de régulation n'entrant pas en contradiction avec les règles du marché. La réglementation bancaire actuelle s'appuie sur la supervision, la discipline du marché et les ratios prudentiels, en particulier les ratios des fonds propres minimaux. L'objet de cet article est la présentation de l'architecture du nouvel accord de Bâle (1999) qui se base sur trois piliers se consolidant mutuellement.

***Mots clés :*** *Banque, Réglementation prudentielle, Approche standard, Approche de notation interne, Risque de crédit, Risque de marché, Risque opérationnel, Surveillance prudentielle, Discipline de marché.*

**Abstract** *:*

In order to adapt to the liberalization of the financial sphere started in the Eighties, marked in particular by the end of the framing of credit, the disappearance of the various forms of protection of the State whose profited the banks, and the privatization of the near total of the establishments in Europe, the banking regulation evolved to a prudential approach, perceived like the only mode of regulation not entering in contradiction with the rules of the market. The current banking regulation is pressed on the supervision, the discipline of the market and the ratios prudential; in particular the ratios of the minimal own capital stocks. The object of this article is the presentation of the architecture of the new agreement of Basle (1999) which is based on three pillars consolidating it self mutually.

---

[1] Doctorant en Sciences Economiques - CEMF-LATEC : Centre d'Etudes Monétaires et Financières - Laboratoire d'Analyse et de Techniques Economiques (CNRS) / Université de Bourgogne (Dijon). Email : Hamza.Fekir@u-bourgogne.fr

# INTRODUCTION :

Plus de dix années se sont écoulées depuis que le Comité de Bâle sur le contrôle bancaire a instauré l'accord sur les fonds propres (l'accord de 1988). Depuis lors, les activités bancaires, les pratiques de gestion des risques, les méthodes prudentielles et les marchés financiers ont connu d'importantes mutations. En juin 1999, le Comité avait publié une série de propositions destinées à remplacer l'accord de 1988 par un dispositif plus différencié en fonction du risque.

L'accord de 1988, reposait sur le montant total des fonds propres par établissement, crucial pour réduire le risque de défaillance pour les déposants. Perfectionnant cette base, le nouveau dispositif vise non seulement à lier plus étroitement les normes de fonds propres au risque effectif mais aussi à renforcer le contrôle et à uniformiser l'information financière avec pour objectif de fond la garantie de la solidité du système bancaire international[2].

Bien que, ce nouveau dispositif s'adresse prioritairement aux grandes banques internationales, ses principes de base sont conçus pour convenir à des établissements présentant des degrés variables de complexité et de technicité. Pour l'élaborer, le Comité a consulté des autorités de contrôle du monde entier et il s'attend qu'à l'issu d'un certain délai les grandes banques, quel que soit leur pays d'origine, y auront adhéré[3].

Le nouveau dispositif (Bâle II) offre une gamme d'options allant de mécanismes simples aux méthodologies avancées pour mesurer le risque de crédit et le risque opérationnel[4] afin de déterminer les niveaux des fonds propres. Il prévoit une architecture souple dans laquelle les banques, dans le cadre du processus de surveillance prudentielle, adopteront l'option la mieux adaptée à leur niveau de technicité et à leur profil de risque. Il introduit aussi expressément des incitations en faveur de mesures du risque plus rigoureuses et plus exactes.

Le nouvel accord est destiné à établir des approches à la fois plus exhaustives et plus différenciées en fonction du risque que l'accord de 1988, tout en préservant le niveau global de fonds propres réglementaires. Des exigences en fonds propres plus conformes aux risques permettront aux banques de gérer leurs activités avec davantage d'efficience.

---

[2] Enjeux du nouveau ratio, (mai 2003), *La fédération bancaire française*.
[3] Vue d'ensemble du Nouvel accord de Bâle sur les fonds propres (Avril 2003), Document soumis à consultation, *Comité de Bâle sur le contrôle bancaire, Banque des règlements internationaux*
[4] Le risque opérationnel constitue une nouveauté dans ce dispositif car avant cet accord ce risque n'a pas été pris en considération.

Le but du présent article, est la présentation sur la base des données actuellement disponibles l'architecture du nouvel accord de Bâle qui se base sur trois piliers se consolidant mutuellement, pour le premier, des exigences minimales en fonds propres, assurer pour le second un processus de surveillance prudentiel et inciter à une discipline de marché pour le dernier.

## 1. LES OBJECTIFS DE LA REFORME PROPOSEE :

Pourquoi le Comité de Bâle s'est-il engagé dans ce projet long et difficile que constitue la réforme du dispositif prudentiel actuel « Ratio Cooke » ?

L'accord de 1988 s'est traduit par une augmentation très significative des fonds propres pour la quasi-totalité des banques internationales. Toutefois, le ratio de solvabilité actuel (Ratio Cooke) ne constitue plus qu'une mesure simplificatrice des risques auxquels une banque est exposée et ne permet pas de prendre en compte les nombreuses évolutions technologiques qui ont eu lieu dans le domaine de la finance, de même que ce ratio est de moins en moins bien adapté aux risques effectivement courus par les banques, ainsi que le régime actuel n'incite pas assez les établissements à gérer ces risques avec prudence. Au pire, ce régime entraîne une mauvaise affectation des ressources, il offre des possibilités d'arbitrage entre les réglementations (autrement dit, il permettrait d'éluder les exigences de fonds propres) et il manque de la souplesse nécessaire pour tenir compte des innovations du marché. Cette possibilité à l'arbitrage réglementaire a compromis l'efficacité de Bâle I, en tant qu'indicateur de la solvabilité d'un établissement de crédit, et a naturellement plaidé en faveur d'un ratio à la fois plus flexible et plus sensible aux risques. C'est la raison pour laquelle, le Comité de Bâle a considéré, à la fin des années quatre-vingt-dix, qu'il ne pouvait pas différer plus longtemps cette réforme[5]. C'est donc afin de remédier à ces insuffisances que le Comité de Bâle a entamé, dès 1998, une réflexion sur la réforme du ratio de solvabilité, avec quatre objectifs principaux :

---

[5] Danièle NOUY., (2003), l'économie du nouveau dispositif et les conséquences de la nouvelle réglementation, *Revue d'économie financière,* Décembre.

- Développer et promouvoir la solidité et la stabilité du système financier avec un ratio de solvabilité qui serait plus sensible aux risques réellement encourus par un établissement ;
- Améliorer les conditions de la concurrence bancaire en éliminant les possibilités d'arbitrage réglementaire ;
- Elaborer une approche plus exhaustive de contrôle des risques bancaires ;
- S'adresser principalement aux banques actives au niveau international, tout en permettant une application à des banques présentant différents niveaux de complexité et de sophistication [6].

Le Nouvel accord a ainsi été conçu avec l'idée qu'il devra, pour le moins, préserver le niveau actuel des fonds propres dans l'ensemble du système bancaire, donc, en moyenne, il ne devrait générer, ni une augmentation nette, ni une diminution nette des fonds propres minimaux.

De plus, le comité de Bâle considère que les exigences de fonds propres peuvent, et doivent, être alignés sur les meilleures pratiques de gestion des établissements de crédit. Avec des exigences en fonds propres plus sensibles aux risques, le capital des banques sera utilisé plus efficacement pour couvrir ces risques, financer l'économie et assurer la stabilité financière.

## 2. LA STRUCTURE GLOBALE DU NOUVEL ACCORD :

Le nouveau dispositif prudentiel repose sur trois piliers se consolidant mutuellement ; ces trois éléments réunis devraient contribuer à la sécurité et à la solidité du système financier. Le Comité insiste sur la nécessité d'une application rigoureuse simultanée des trois piliers et entend coopérer activement avec les autorités de contrôle bancaire pour parvenir à une mise en œuvre efficace de tout aspect de ce dispositif. Seule une mise en œuvre concomitante et équilibrée des trois piliers constitue une application correcte de Bâle II susceptible de produire à terme tous les effets positifs escomptés[7].

---

[6] Le Nouvel accord de Bâle sur les fonds propres, (exercice 2001), Conseil national du crédit et du titre.
[7] Danièle NOUY., (2003), l'économie du nouveau dispositif et les conséquences de la nouvelle réglementation, Revue d'économie financière, Décembre

## 2.1 Premier pilier : Exigences minimales des fonds propres :

Les mesures quantitatives concernant les exigences de fonds propres constituent le point de départ du nouvel accord, de même que pour la première fois, le Comité de Bâle stipule des exigences de fonds propres explicites en regard du risque opérationnel. Si les réglementations concernant les risques du marché n'ont pas changé, il n'en est pas de même pour les procédures de mesure du risque de crédit qui elles sont plus élaborées que dans l'accord de 1988. Le Comité de Bâle propose une approche évolutive, permettant aux banques de choisir, sous le contrôle de leurs superviseurs la méthode la mieux adaptée à leur profil de risque et au degré de sophistication de leurs outils de gestion interne. Dans le nouveau dispositif, la définition des fonds propres et l'exigence minimale de 8 % restent les mêmes, mais des pourcentages ont été définis afin d'affecter le montant des fonds propres aux différents risques (75 % des fonds propres seront affectés au risque de crédit ce qui correspond à une équivalence de 6 %, 20 % au risque opérationnel et 5 % au risque du marché).

L'objectif de ce premier pilier est d'améliorer le calcul des risques et leur couverture par les fonds propres afin d'assurer une meilleure stabilité micro-prudentielle avec un ratio mieux proportionné aux risques[8]. Afin de mieux comprendre la logique qui le sous-tend, nous allons étudier au préalable les deux termes du ratio de solvabilité (Mc Donough) à savoir : Les fonds propres et les risques qui constituent, respectivement le numérateur et le dénominateur du ratio.

### 2.1.1 Pourquoi les fonds propres ont autant d'importance ?

Pourquoi le Comité de Bâle impose aux banques de disposer d'un montant minimum de fonds propres proportionnel à leur risques ?

Afin de mieux comprendre l'utilité des fonds propres, on va suivre une démarche qui consiste à passer de leur rôle (fonds propres) dans le cadre d'une entreprise à une application au cas d'une banque.

---

[8] PIERRE-YVES Thoraval, ALAIN Duchateau., Stabilité financière et nouvel accord de Bâle, *Secrétariat général de la Commission bancaire.*

### 2.1.1.1 *Application au cas d'une entreprise :*

On peut dire que le bilan d'une entreprise se décompose schématiquement de la façon suivante :

Au passif figurent les sources de financement :

- Capitaux.
- Dettes à plus ou moins long terme.

A l'actif figure tout ce que l'entreprise a réalisé grâce aux financements apportés, tout ce qu'elle possède :

- Les immobilisations corporelles (immeubles) et incorporelles (participation dans d'autres sociétés).
- Les stocks.
- Les créances.

Le bilan est présenté de telle sorte que le total l'actif = total passif

Les capitaux propres sont l'ensemble des ressources « couvrant le risque » de l'entreprise, c'est-à-dire celles qui ne seront en principe remboursées qu'avec la liquidation de l'entreprise (fonds propres), ou celles qui ne doivent être remboursées qu'à très longue échéance (quasi-fonds propres).

L'actif net est quant à lui égal à l'ensemble des avoirs de l'entreprise diminué de l'ensemble de ses engagements réels ou potentiels :

Actif net = actif immobilisé + actif circulant et financier – ensemble des dettes.

La solvabilité d'une entreprise, quelle qu'elle soit, est sa capacité à rembourser l'intégralité de ses engagements en cas de liquidation totale. Elle dépend donc de la qualité de ses actifs et plus particulièrement de la facilité avec laquelle ceux-ci peuvent être liquidés et du montant de ses engagements.

Comme actif = passif

Capitaux propres + dettes = Actif immobilisé + Actif circulant et financier

Capitaux propres = actif net.

Conclusion : La solvabilité qui intuitivement correspond au rapport Dettes / Actif net, peut également se mesurer par le rapport Dettes / Capitaux propres.

*2.1.1.2 Application au cas d'une banque :*

Généralement, on reconnaît la solvabilité de la banque par sa capacité à faire face aux demandes de retrait de ses déposants, et cela fait partie de la responsabilité des autorités de tutelle, de s'assurer que les banques sont bien aptes à faire face à leurs obligations. Il y va en effet de la stabilité de l'économie tout en entière d'un pays.

Pour une banque, les dettes sont essentiellement constituées des dépôts à vue. Les actifs financiers sont constitués des crédits octroyés ; c'est en effet la finalité d'une banque de distribuer du crédit.

A partir de l'égalité vue plus haut, on voit que la banque puisse distribuer davantage de crédit, elle doit soit collecter davantage de dépôts au risque de ne pas pouvoir rembourser ceux-ci, soit renforcer ses capitaux propres.

Or, une entreprise se trouve davantage en sécurité si une partie de son actif circulant n'est pas financée par des ressources qui viendront à échéance dans l'année. L'actif présente toujours un caractère aléatoire et donc risqué ; en particulier quand il est constitué essentiellement de créances comme pour les banques, alors que les dettes elles sont inéluctables ; c'est pour quoi il faut qu'une partie de l'actif soit financée non pas par les dettes mais par du capital.

D'autre part, si on impose à une banque d'augmenter ses fonds propres elle a plus à perdre en cas de faillite et aura donc tendance à adopter des activités moins risquées. Donc, le niveau des fonds propres est garant de la solidité financière de l'entreprise. Les fonds propres sont donc garants de la solvabilité de la banque face aux pertes que les risques pris à l'actif sont susceptibles d'engendrer et influencent aussi sa rentabilité et ses incitations à la prise de risques.

Pour toutes ces raisons, le ratio de solvabilité dans le cas des banques s'exprimait initialement par le rapport du montant des fonds propres au montant des crédits distribués, ceux-ci étant pondérés par leur caractère plus au moins risqué.

Après avoir présenté l'importance des fonds propres, nous allons maintenant étudier la façon par laquelle le Comité de Bâle suggère de comptabiliser les différents postes de l'actif et hors bilan constitutifs du risque de crédit, du risque de marché et du risque opérationnel.

*2.1.2 Risque de crédit :*

Généralement le risque de crédit est défini par le risque qu'un débiteur fasse défaut ou que sa situation économique se dégrade au point de dévaluer la créance que l'établissement détient sur lui. Pour la mesure de ce type de risque, le Comité de Bâle propose deux grandes options qui sont : l'approche standard et l'approche fondée sur les notations internes[9] (International Rating Based approche ou IRB). Cette dernière comporte deux variantes, simple et avancée.

*2.1.2.1 L'approche standard :*

Bien qu'identique dans son principe à l'accord de 1988, le nouvel accord redéfinit la pondération des risques des différents actifs et des positions hors bilan afin de la rendre beaucoup plus sensible aux risques. La pondération de chaque actif est fonction du rating attribué par des agences de notation (comme Standard & poor's, Moody's) ou autres organismes (Banque de France), les coefficients de pondération sont fixés par grande catégorie d'emprunteurs (souverain, banque ou entreprise).

L'approche standard constitue une méthode simple accessible à toute banque, le principe du calcul du risque de crédit est le suivant : à chaque actif ou élément hors bilan se verra affecter un coefficient variable de 0 % à 100 %, selon le risque qu'il présente. Si des notations externes des débiteurs ne sont pas disponibles, des pondérations forfaitaires sont prévues, par exemple 100 % pour les actifs non cotés donc la même actuellement, de même que pour certains types d'actifs particulièrement risqués sur des banques ou des entreprises ayant déjà connu des défaillances pourrant être pondérés à plus de 100 %.

---

[9] il faut noter que le recours à l'approche IRB sera soumis à l'agrément des autorités de contrôle sur la base de critère définis par le comité de Bâle.

**Tableau 01 :** Pondérations proposées par le Comité de Bâle par nature de contrepartie et par note.

|  | AAA à AA- | A+ à A- | BBB+ à BBB- | BB+ à BB- | B+ à B- | < B- | Non coté |
|---|---|---|---|---|---|---|---|
| Souverain | 0 % | 20 % | 50 % | 100 % | 100 % | 150 % | 100 % |
| Banque | 20 % | 50 % | 50 à 100 % | 100 % | 100 % | 150 % | 50 à 100 % |
| Banque – actif à CT | 20 % | 20 % | 20 % | 50 % | 50 % | 150 % | 20 % |
| Entreprise | 20 % | 50 % | 100 % | 100 % | 150 % | 150 % | 100 % |

**Source :** Le Comité de Bâle, Banque des Règlements Internationaux, texte de Janvier 2001.

La lecture du tableau est la suivante : les risques des crédits à la catégorie de souverain des pays les mieux notés (de AAA à AA-) n'auraient pas à être provisionnés tandis que ceux des pays les moins bien notés (sous B-) devraient être provisionnés à hauteur de 12 % (150 % de 8 %).

Donc, le calcul à partir du bilan s'effectue en appliquant aux différents natures d'engagements un des coefficients de pondération cités précédemment. Le montant du risque pondéré des engagements s'obtient par la formule suivante :

$$\text{Engagements au bilan} \times \text{Taux de pondération du risque} = \text{Risque pondéré.}$$

Donc, les différents postes de l'actif, selon la nature de la contrepartie, verront leur montant multiplié par le coefficient de risque correspondant. La procédure est la même pour les éléments hors bilan[10], à la différence tout de même que ces différents éléments seront auparavant pondérés par des coefficients de conversion, fonction des catégories hors bilan.

---

[10] Parmi les éléments hors bilan, on note : Obligations cautionnées, Crédits documentaires ou les marchandises servent de garantie, Cautions, avals et autres garanties accordés à la clientèle ou à des établissements de crédit.

Au total, la banque devra additionner les éléments de l'actif, les éléments hors bilan, le tout pondéré par les coefficients de risque. Elle obtiendra ainsi la valeur de l'actif fonction du risque de contrepartie.

L'exemple suivant va nous permettre de mieux comprendre l'opération de pondération du risque d'un élément hors bilan:

Soit l'opération d'ouverture de crédit moyen terme confirmée accordée à une collectivité locale d'un pays membre de l'OCDE pour un montant de 10 millions euros :

Le risque pondéré serait de : 10 Millions euros $\times$ 50 % $\times$ 20 % = 1 million euros.

On note que :
- 50 % représente le taux de conversion.
- 20 % représente coefficient de risque qui est déterminé en fonction de la nature de l'opération et de l'emprunteur ;

Toutefois, il faut noter que le Comité de Bâle laisse le choix des méthodes aux banques, la méthode standard relativement simple à mettre en place, a vocation à être utilisée que par les banques n'ayant pas les moyens techniques et humains d'utiliser les méthodes fondées sur les notations internes. Cette méthode accorde un rôle clef aux agences de notation, ce qui constitue l'une des principales critiques adressées par les professionnels. En effet, le texte proposé laisse intervenir dans la réglementation bancaire des organismes privés « les agences de notation » répondant à des objectifs de recherche de rentabilité et non de service public.

Ces objectifs peuvent être contradictoires ; de plus une autre critique est avancée, celle des erreurs commises par les agences de notation. Mais, on peut opposer à cet argument, la constatation que les erreurs commises par les agences demeurent marginales au regard de la taille du portefeuille de notes qu'elles suivent.

*2.1.2.2 La méthode de notation interne :*

Dans cette approche les banques pourront utiliser leurs estimations internes sur la solvabilité de leurs emprunteurs pour évaluer le risque de crédit inhérent à leur portefeuille, à condition qu'elles respectent des critères stricts en matière de méthodologie et de communication financière[11].

Les banques concernées par ces approches s'appuient sur leurs estimations internes des composantes du risque pour déterminer l'exigence des fonds propres en regard d'une exposition donnée. Ces approches internes permettent d'obtenir un rating et une probabilité de défaut pour chaque actif. Les pondérations des actifs sont déterminées en fonction de quatre variables : la probabilité de défaut des emprunteurs, le taux de recouvrement des pertes, le montant exposé au risque et la maturité des engagements.

$$RW = f(PD, LGD, EAD, M)$$

Avec: RW = Pondération de risque (R*isk Weight*).
PD = Probabilité de défaut (*Probability of Default*).
LGD = 1-taux de recouvrement (*Loss Given Default*).
EAD = Montant en risque (*Exposure At Default*).
M = Maturité de l'actif (*Maturity*).

Le dispositif prévoit deux méthodologies pour les prêts aux entreprises, aux emprunteurs souverains et aux banques, la première est une évaluation *simplifiée* dans laquelle les banques estiment elles-mêmes la probabilité de défaut de chaque client et les régulateurs fournissent les autres éléments d'appréciation (taux de recouvrement de 50 %, exposition au risque égale à la valeur nominale des actifs, et maturité de 3 ans), la deuxième méthode est dite *avancée,* a priori, destinée aux seuls grands établissements dotés de systèmes sophistiqués d'allocation du capital et calculant eux-mêmes l'ensemble des paramètres nécessaires. Il s'agit d'une structure incitative, car l'exigence en fonds propres en approche avancée sera plus faible qu'en approche de base (fondation).

---

[11] Eric- PAGET BLANC., Le rôle informationnel des ratios de fonds propres des banques, *Document de travail, université d'Evry- Val D'Essonne.*

*2.1.3 Risque de marché :*

Le risque de marché est le risque de perte ou de dévaluation sur les positions prises suites à des variations des prix (cours, taux) sur le marché. Ce risque s'applique aux instruments suivants : produits de taux (obligations, dérivés de taux), actions, change, matières premières.

Le nouveau dispositif n'apporte pas des grands changements pour le risque de marché, on note la prise en compte des instruments de réduction des risques, tels que les sûretés financières, les garanties, la compensation et on trouvera toujours les deux méthodes d'évaluation de risque suivantes :

- Une méthode standard.
- Une approche modèle interne (VaR).

*2.1.4 Risque opérationnel :*

Des exigences minimales de fonds propres sont aussi prévues pour couvrir le risque opérationnel, défini comme un « risque de pertes directes ou indirectes d'une inadéquation ou d'une défaillance attribuable à des procédures, personnes, systèmes internes ou à des événements extérieurs, par exemple : un contrat mal rédigé ou bien une défaillance informatique[12] ». La définition inclut le risque juridique, mais exclut les risques stratégiques et d'atteinte à la réputation.

Pour la mesure du risque opérationnel, le Comité de Bâle, présente trois méthodes de calcul des exigences de fonds propres en regard de ce type de risque, par ordre croissant de complexité et de sensibilité au risque : approche indicatrice de base, approche standard et approche de mesures complexes (AMC)[13].

---

[12] Nouvel accord de Bâle sur les fonds propres (Avril 2003), Document soumis à consultation, article 607, *Comité de Bâle sur le contrôle bancaire, Banque des règlements internationaux*.
[13] Nouvel accord de Bâle sur les fonds propres (Avril 2003), Document soumis à consultation, article 608, *Comité de Bâle sur le contrôle bancaire, Banque des règlements internationaux*.

Les banques sont invitées à passer de l'approche la plus simple à la plus complexe à mesure qu'elles développent des systèmes et des pratiques de mesure plus élaborés du risque opérationnel.

Le Comité donne la possibilité aux banques d'utiliser l'approche indicatrice de base ou l'approche standard pour certaines parties de ses activités et (AMC) pour d'autres, à condition de satisfaire à certains critères minima[14].

De même qu'un établissement ne pourra pas sans l'approbation de l'autorité de contrôle, revenir à une approche plus simple après avoir été autorisé à utiliser une approche plus élaborée. En outre, si une autorité détermine qu'une banque ne répond plus aux critères d'agrément pour une approche, elle peut lui demander de retourner à une approche plus simple pour une partie ou l'ensemble de ses activités, jusqu'à ce qu'elle satisfasse aux conditions posées par l'autorité de contrôle pour utiliser à nouveau une approche plus élaborée. Quoi qu'il en soit, l'utilisation des méthodes plus complexes ne pourra se faire par la banque qu'en cas du respect des deux éléments suivants :

- Des saines pratiques élaborées pour déceler, surveiller et contrôler le risque opérationnel ;
- Des obligations de communication d'informations qualitatives et quantitatives concernant leurs méthodes de calcul de fonds propres les processus internes de gestion et de contrôle des risques opérationnels.

### 2.1.4.1 Approche indicatrice de base :

Les banques appliquant cette approche doivent, en regard du risque opérationnel, détenir un montant des fonds propres correspondant à un pourcentage fixe (alpha) de leur produit net bancaire moyen sur les trois dernières années[15]. L'exigence peut être exprimée ainsi :

$$KNI = GI \times \alpha$$

Où :

---

[14] Nouvel accord de Bâle sur les fonds propres (Avril 2003), Document soumis à consultation, article 640 et 641, *Comité de Bâle sur le contrôle bancaire, Banque des règlements internationaux*.
[15] Nouvel accord de Bâle sur les fonds propres (note explicative) avril 2003, *Secrétariat du Comité de Bâle sur le contrôle bancaire, Banque des règlements internationaux*.

KNI = Exigence en fonds propres dans l'approche indicateur de base ;
GI = Produit annuel brut moyen sur les trois dernières années ;
α = 15 %, valeur, fixée par le Comité, représentant la relation entre l'exigence en fonds propres pour l'ensemble du secteur et l'indicateur pour l'ensemble du secteur.

Dans l'article 613 du même dispositif, le Comité de Bâle défini le produit net bancaire par l'ensemble des intérêts créditeurs nets et autres produits d'exploitation. Il exclut les éléments suivants : provisions (pour intérêts impayés, par exemple) ; plus ou moins-values matérialisées en liaison avec la cession de titres du portefeuille bancaire ; les éléments extraordinaires ou inhabituels et produits des activités d'assurance.

### 2.1.4.2 Approche standard :

Pour l'utilisation de l'approche standard, le Comité de Bâle dans son dispositif relatif au nouvel accord de Bâle sur les fonds propres (avril 2003) réparti les activités des banques en huit catégories ou lignes de métiers : financement des entreprises, négociation et vente, banque de détail, banque commerciale, paiement et règlement, fonctions d'agent, gestion d'actifs et courtage de détail.

Pour chaque catégorie, le produit brut sert d'indicateur global approché du volume et, partant, du degré d'exposition au risque opérationnel ; l'exigence en fonds propres est calculée en multipliant le produit brut par un facteur (bêta) spécifique. Bêta représente une mesure approchée de la relation, pour l'ensemble du secteur bancaire, entre l'historique des pertes imputables au risque opérationnel pour une catégorie donnée et le montant agrégé du produit brut de cette catégorie d'activité. Il convient de noter que dans l'approche standard, le produit brut se mesure par catégorie et non pour l'ensemble de l'établissement ; s'agissant du financement des entreprises, par exemple, l'indicateur est le produit brut qui lui est spécifique.

Pour le calcul du produit brut, on exclut les plus ou moins-values sur titres classés comme « détenus jusqu'à échéance » et « disponibles à la vente », qui sont des éléments courants du portefeuille bancaire. L'exigence totale en fonds propres représente la somme des exigences en fonds propres pour chacune des catégories d'activité. Elle peut être exprimée ainsi :

$$KTSA = \sum(GI_{1-b} \times \beta_{1-b})$$

Où :

KTSA = Exigence en fonds propres selon l'approche standardisée,

$GI_{1-b}$ = Produit annuel brut moyen sur les trois dernières années, tel que défini dans l'approche indicateur de base, pour chacune des huit catégories,

$\beta_{1-b}$ = Pourcentage fixe, déterminé par le Comité, représentant la relation entre le niveau en fonds propres requis et le produit brut de chacun des huit catégories.

**Tableau 02** : Les valeurs bêta utilisées dans l'approche standard.

| *Activité* | *Catégories d'activité* | *Mesure de l'activité* | *Facteur bêta ($\beta$)* |
|---|---|---|---|
| Banque d'investissement | Financement des entreprises | Revenu brut | 18 % |
|  | Négociation et vente | Revenu brut | 18 % |
| Banque | Banque de détail | Actifs moyens | 12 % |
|  | Banque commerciale | Actifs moyens | 15 % |
|  | Paiement et règlement livraison | Volume | 18 % |
| Autres | Fonctions d'agent | Revenu brut | 15 % |
|  | Gestion d'actifs | Revenu brut | 12 % |
|  | Courtage de détail | Actifs et conservation | 12 % |

**Source :** Nouvel accord de Bâle sur les fonds propres (note explicative), avril 2003, *Secrétariat du Comité de Bâle sur le contrôle bancaire, Banque des règlements internationaux.*

Cette approche est donc plus sensible au risque mais suppose une estimation du risque relatif de chaque ligne de métier.

### 2.1.4.3 *Approches de mesures complexes (AMC) :*

Dans les articles 618 et 619 du même dispositif relatif au nouvel accord sur les fonds propres (avril 2003), on constate que ces méthodes de mesures complexes sont basées sur des approches internes fondées sur la modélisation, par chaque établissement, de la distribution de pertes à partir de données internes ou externes ou encore d'analyse de scénarios. Conceptuellement séduisante car très sensible au risque, cette approche est néanmoins techniquement difficile à mettre en œuvre en raison d'historiques de données encore faibles. En tout état de cause, elle ne pourra être utilisée que par les établissements dont les systèmes de gestion du risque opérationnel et de collecte des données auront été validés par l'autorité de contrôle, de même que la banque qui adoptera ces approches de mesures complexes doit calculer ses exigences en fonds propres grâce à cette méthodologie pendant une année avant la mise en œuvre du Nouvel accord fin 2006.

Le dispositif incite à opter pour la méthode avancée, celle-ci étant en principe moins consommatrice en fonds propres réglementaires. En retour, l'économie se « paye » par la mise en place d'une organisation spécifique visant à un meilleur contrôle des risques opérationnels, et en définitive, à la réduction des pertes. Ainsi, contrairement à l'approche de base, l'approche standard impose que soient identifiés et évalués les risques opérationnels. L'approche avancée requiert quant à elle la nomination d'une entité indépendante responsable de la mise en place d'une stratégie de réduction des risques opérationnels.

### *2..1.5* Le calcul du ratio final :

Pour préserver la cohérence du calcul, les montants de fonds propres requis au titre de risque de marché et du risque opérationnel doivent être multipliés par 12,5 (l'inverse de 8 %) avant de les incorporer au calcul final.

Risque de crédit = Actifs pondérés en fonction de leur risque.
Risque de marché = Capital requis pour la couverture du risque de marché × 12,5.
Risque opérationnel = Capital requis pour la couverture du risque opérationnel × 12,5.

$$\text{Ratio Mc Donough} = \frac{\text{Total des fonds propres}}{\text{Risque de crédit + Risque de marché + Risque opérationnel}} \geq 8\%$$

## 2.2 Deuxième pilier : Processus de surveillance prudentielle :

Le deuxième pilier du nouveau dispositif vise à introduire davantage de cohérence entre les risques pris par une banque et l'allocation des fonds propres au sein de cette dernière, ce pilier repose sur quatre principes fondamentaux :

- Les banques doivent disposer d'un système de mesure interne de l'adéquation de leur fonds propres à leur profil de risques et d'une stratégie de maintien de cette adéquation ;
- Les autorités de contrôle doivent examiner ce système de mesure et cette stratégie est s'assurer de leur conformité avec la réglementation ;
- Les autorités de contrôle attendent des banques qu'elles disposent de fonds propres supérieurs à ceux fixés réglementairement et doivent pouvoir le leur imposer ;
- Les autorités de contrôle doivent pouvoir intervenir de manière préventive afin d'éviter que les fonds propres des banques ne tombent en deçà de niveaux prudents et doivent pouvoir leur imposer une action correctrice si le niveau de ces derniers n'est pas maintenu ou restauré[16].

Donc, le processus de surveillance constitue un complément essentiel aux mesures de fonds propres réglementaires et aux règles générales définies par le premier pilier. Il permet de vérifier l'adéquation des fonds propres de la banque sur la base de l'évaluation complète des risques qu'elle encourt. Une fois que les autorités de contrôle ont vérifié les procédures internes d'affectation de fonds propres réalisées par la banque, elles peuvent revoir à la hausse les exigences minimales des fonds propres.

## 2.3 Troisième pilier : Discipline de marché.

Depuis quelques années déjà, les autorités de contrôle considèrent que la qualité de l'information financière est un élément fondamental de l'efficience des marchés et de la solidité des systèmes financiers.

---

[16] Nouvel accord de Bâle sur les fonds propres (note explicative), avril 2003, *Secrétariat du Comité de Bâle sur le contrôle bancaire, Banque des règlements internationaux.*

En s'inspirant de ses recommandations antérieures, le Comité de Bâle a défini un ensemble d'informations, que les banques devront publier sur un rythme semestriel, par exemple : touchant au champ d'application du ratio (consolidation), le niveau et la structure détaillée des fonds propres ou même l'exposition au risque/mode de gestion de risque (crédit, marché, opérationnel, taux ….).

La logique qui sous-tend le troisième pilier est que l'amélioration de la communication financière permet de renforcer la discipline de marché, perçue comme un complément à l'action des autorités de contrôle. L'information financière est, en effet, toujours une incitation à rationaliser la gestion des risques pour traduire la nécessaire cohérence dans la démarche des banques entre leur système de gestion interne, de même qu'en communiquant des informations détaillées sur tous les types de risque, une banque permet à tous les autres acteurs du marché de mieux analyser son profil de risque et l'adéquation de ses fonds propres, de même que l'utilisation des méthodes avancées sera conditionnée par la publication de ces informations. Cette préoccupation rejoint celle de la transparence financière[17].

## 3. Comparaison entre l'accord de Bâle I et l'accord de Bâle II :

L'accord de 1988 ne pose que le principe d'une exigence quantitative fondée sur une méthode de calcul uniforme. Le futur dispositif reposera sur trois types d'obligations (les piliers) :

- Les établissements devront disposer d'un montant de fonds propres au moins égal à un niveau calculé selon l'une des méthodes proposées (Pilier I) ;
- Les autorités disposeront de pouvoirs renforcés et pourront en particulier imposer, au cas par cas, des exigences supérieures à celles résultant de la méthode utilisée (Pilier II) ;
- Les établissements seront soumis à la discipline de marché (Pilier III), étant tenus de publier des informations très complètes sur la nature, le volume et les méthodes de gestion de leurs risques ainsi que sur l'adéquation de leurs fonds propres.

Par rapport au dispositif actuel, le futur Accord comporte cinq novations principales :
- Des exigences en fonds propres s'imposeront non seulement pour les risques de crédit et pour les risques de marché mais aussi pour les risques opérationnels ;

---
[17] CHRISTIAN NOYER., Bâle II : Genèse et enjeux, Conférence- débat, *association d'économie financière.*

- Pour calculer les exigences en fonds propres au titre de chaque type de risque, les établissements se verront ouvrir plusieurs options, notamment entre des méthodes standards et des méthodes fondées sur des notations ou des mesures internes ;

Le mode de calcul de ces exigences intégrera davantage la réalité des risques, notamment par une meilleure prise en compte des techniques de réduction des risques ;

- Les exigences en fonds propres pourront être adaptées individuellement en fonction du profil de risque de chaque établissement, les autorités de contrôle pouvant imposer des exigences individuelles supérieures à celles calculées dans le cadre du pilier 1 ;
- Les établissements devront publier des informations détaillées sur leurs risques et l'adéquation de leurs fonds propres.

## *Bâle I :*

$$\text{Ratio Cook} = \cfrac{\text{Fonds propres}}{\left[\begin{array}{c}\textit{\textbf{Risque de crédit :}} \\ \text{Assiette du risque} \\ \text{est mesurée par :} \\ \bullet \text{ Approche standard}\end{array}\right] + \left[\begin{array}{c}\textit{\textbf{Capital requis pour la couverture du risque de marché :}} \\ \text{Assiette du risque est mesurée par :} \\ \bullet \text{ Approche standard} \\ \bullet \text{ Approche de notation interne}\end{array}\right] \times 12{,}5} \geq 8\,\%$$

**Bâle II :** (* Représente tous ce qui est nouveau par rapport à Bâle I)

*Pilier 01 :* *Exigences minimales de fonds propres.*

$$\text{Ratio Mc Donough} = \cfrac{\text{Fonds propres}}{\begin{array}{c}\left[\begin{array}{c}\textit{\textbf{Risque de crédit :}} \\ \text{Assiette du risque est mesurée par :} \\ \bullet \text{ Approche standard (modifiée)} \\ \bullet \text{ Approche de notation interne de base*} \\ \bullet \text{ Approche de notation interne avancée*}\end{array}\right] + \left[\begin{array}{c}\textit{\textbf{Capital requis pour la couverture du risque de marché :}} \\ \text{Assiette du risque est mesurée par :} \\ \bullet \text{ Approche standard} \\ \bullet \text{ Approche de notation interne.}\end{array}\right] \times 12{,}5 \\ + \left[\begin{array}{c}\textit{\textbf{Capital requis pour la couverture du risque opérationnel :}} \\ \text{Assiette du risque est mesurée par :} \\ \bullet \text{ Approche standard*} \\ \bullet \text{ Approche indicatrice de base *.} \\ \bullet \text{ Approche de mesures complexes*.}\end{array}\right] \times 12{,}5\end{array}} \geq 8\%$$

*Pilier 02* : Processus de surveillance prudentielle*.

*Pilier 03 :* Recours à la discipline de marché, via une communication financière efficace*.

**CONCLUSION :**

Bâle II est tout d'abord un dispositif plus complet que Bâle I dans la mesure où l'ensemble des risques auxquels est exposée une banque devront être pris en compte dans l'appréhension du profil de risque de cette dernière. L'introduction d'une exigence de couverture du risque opérationnel par des fonds propres, au titre du pilier 1, n'en est pas la seule illustration. Le nouveau dispositif est plus flexible, et plus sensible aux risques que la norme actuelle. L'approche proposée par le Comité de Bâle ne pouvait être qu'évolutive dès lors que l'ambition de la réforme était de promouvoir l'adoption des meilleures pratiques de la profession. C'est pourquoi le Comité a développé pour chaque type de risque, à l'instar, toutes proportions gardées, du régime existant en matière de risques de marché, un menu d'options en tenant compte de l'état de l'art au sein des établissements. Le pilier 1 du nouveau dispositif offre ainsi plusieurs options aux banques pour calculer les exigences des fonds propres relatives à leur risque de crédit et à leur risque opérationnel, chacune de ces banques ayant le choix de retenir l'option la mieux adaptée à son degré de sophistication et à son profil de risque. La partie la plus innovante de Bâle II est constituée à cet égard, comme cela a été abondamment rappelé, par la possibilité offerte aux banques d'utiliser, dans les limites fixées par le nouveau cadre et sous le contrôle de leur autorité de tutelle, leurs propres systèmes internes d'évaluation de leurs risques.

Enfin, la dimension prospective du nouveau dispositif mérite d'être soulignée. L'ambition de ce dernier est en effet d'inciter les banques à mesurer et à gérer leurs risques non seulement de manière plus fine mais aussi de manière plus dynamique. Ainsi, en matière de risque de crédit, le calcul des exigences en fonds propres, au titre du pilier 1, sera effectué par les principaux groupes bancaires à partir de leurs notations internes telles qu'appréciées en principe sur un horizon d'un an mais en pratique sur un horizon plus lointain.

Dans le cadre du pilier 2, les autorités de contrôle sont dotées d'un pouvoir d'appréciation des modèles et pourront imposer aux établissements des exigences individuelles en fonds propres supérieures, de même que ces établissements devront publier des informations détaillées sur leurs risques et l'adéquation de leurs fonds propres (pilier 3).

# Bibliographie :